# Bio-inspired Methods for Dynamic Network Analysis in Science Mapping

*Sándor Soós[1,3] and George Kampis[2,3]*

[1]Institute of Research Organization, Hungarian Academy of Sciences
[2]History and Philosophy of Science, Lorand Eötvös University, Budapest, Hungary
[3]Collegium Budapest, The Institute of Advanced Study
ssoos@colbud.hu, gk@hps.elte.hu

***Abstract***

We apply bio-inspired methods for the analysis of different dynamic bibliometric networks (linking papers by citation, authors, and keywords, respectively). Biological species are clusters of individuals defined by widely different criteria and in the biological perspective it is natural to (1) use different categorizations on the same entities (2) to compare the different categorizations and to analyze the dissimilarities, especially as they change over time. We employ the same methodology to comparisons of bibliometric classifications. We constructed them as analogs of three species concepts: cladistic or lineage based, similarity based, and "biological species" (based on co-reproductive ability). We use the Rand and Jaccard indexes to compare classifications in different time intervals. The experiment is aimed to address the classic problem of science mapping, as to what extent the various techniques based on different bibliometric indicators, such as citations, keywords or authors are able to detect convergent structures in the litrerature, that is, to identify coherent specialities or research directions and their dynamics.

### *1. Introduction: Biological inspiration for the study of the dynamics of science*

Understanding how information propagates in networks is of crucial importance in different contexts ranging from epidemics to the mapping of science. Networks usually have different properties in a time global or time local perspective, and the latter is, in a fundamental sense, "more real" (past or future contacts don't infect, for instance). In the scientometric context, co-author networks and citations networks are seen as dynamically changing over time, forming new nodes as well as links and removing others. To understand this dynamics, the role of the instantaneous network is primary. As a consequence, the formation, existence and dissolution of various scientific aggregates, such the emergence and persistence of concepts, topics, etc. are best mapped in a time stamped series of networks. In this paper, we deal with dynamic clusters of scientific publications to deal with related phenomena.

Our treatment of dynamic clusters of papers is biologically motivated and conceives clusters as analogs of species. We show how different species concepts allow for a different identification and comparison of time dependent communities. Of the many available species concepts (Mayden, 1997) we select three: the biological (relation or link-based), the cladistic (lineage based), and the phenetic (metric clustering) concept. In biology, comparisons of different species (i.e. cluster or community) concepts counts as daily routine and applying (*mutatis mutandis*) the species ideas to scientometric problems can bring forth the possibility of using competitive conceptualizations and tested methodology for hitherto neglected scientometric problems – such as the parallel and dynamic characterization of different dynamic clusters.

In an earlier study, using Artificial Life methods and biological datasets generated by simulation studies (Kampis et al. 2007) we found that in periods of near stasis the three above mentioned methods of biological species identification tended to yield similar (i.e. mutually consistent) classifications - we used the Rand index (Rand, 1971) to characterize this in the quantitative sense. In contrast, in the transition periods (intervals of speciation, viz. intensive cluster dynamics) highly different species classifications (characterized by low Rand indices) were obtained. We may expect similar behavior in the scientometric domain and thus we selected a dataset that might suit dynamic studies.

## *2. Scientometric dimensions*

Deploying different conceptualizations of the species concept in the study of research dynamics can have a methodological relevance in scientometrics. A long-standing issue in scientometrics and science mapping concerns the appropriate method of tracking the evolution of research areas.

Based on the utilization of scholarly and, mostly, bibliographic databases, two general frameworks have emerged: the first is „bibliometrics”, while the second is „automated content analysis”. Bibliometrics, in the narrow sense, focuses on citation patterns to reveal trends from a given body of scientific literature. Content analysis, in contrast, is based on the retrieval of textual characteristics or descriptors of the papers, typically keywords. These descriptors are used as proxies to establish the thematic structure or dynamics of a research field. The two frameworks differ in at least two aspects: the source of information (citations vs. textual descriptors) and the relevant relation for defining the structure of the literature (scientific „species" and their evolution).

In bibliometrics, the type of the latter relation is some kind of actual and „intellectual" kinship realized in citations or co-citations, while in content analysis it is a similarity relation between textual patterns. Due to these differences, the analyses yielded by applying the two methods tend to lead to differing results, even when conducted on the very same corpus or database. The differences give way to a set of questions on

(1) what kind of trends the two methods can extract from the literature,
(2) to what extent the results converge and, in more generally,
(3) what is the relation between the results (whether they stand in a substitutional, complementary or contradictory relation to each other).

The questions (1)-(3) are often addressed in the scientometric literature (cf. King 1987, Braam–Moed–Raan 1991a-b, Noyons–Moed–Luwel 1999).

As can be shown based on the discussion in the first section, this scientometric issue is a reformulation of, or may be conceived in terms of the species problem. The key observation is that the bibliometric model of research fields is analogous to the lineage concept(s) of species, while the content analytic model is on a par with the phenotype or morphological concept(s).

This claim follows from the main feature of the two models as highlighted above: the bibliometric approach draws on an „intellectual" *descendancy* relation between documents, while the content analytic approach focuses on the *similarity* of publications. Contrasting and comparing the two models virtually entails the same set of questions in both the scientometric and the biological uses: these are the questions (1)–(3) indicated in the previous section. As a generalization, these questions can be reformulated as concerning the relationship between the descendancy definition and the similarity definition of scientific and biological species, respectively, in science the descedancy relation being related to research fields/directions and similarity being related to coherent units of scientific activity. Upon this mapping between the species problem and the scientometric issues, and studying the evolution of research fields from the former, the biological perspective may shed light to the latter, which is at the core of scientometric research.

### *3. The experiment*

To investigate the relation between „concepts of scientific species" in different scientometric models, we have designed an experiment addressing primarily question (2) , concerning the degree of convergence between the different models. The main goal of the exercise was to compare and contrast these methodologies on the same dataset, that is, to measure quantitatively the extent to which the trends extracted by the scientometric models were similar or different. The comparative application of these models, as the ususal notion of a trend suggests, involved the pairwise comparison of the dynamics (and evolution) of the research field that the models were capable of tracking.
The scope of our analysis, as the analogy with the species problem imposes, includes three different scientometric models. Two of those were introduced in the previous section already: the bibliometric model as the counterpart of the lineage concept of species, and the content analytic model as the counterpart of the morphological species concept. The third model is based on a sociological conceptualization of research fields, where relations between authors (co-authorsip patterns) are used to circumscribe a speciality or field. This relation, as argued in section 1, can be conceived of as a kind of „intellectual mating", resulting in a joint publication (or „offspring"), which naturally leads to the „biological species concept" (for the notion see section 1). In other words, the third methodology is a social network model of science, by which the list of major species concepts applied to scientometrics also becomes complete.

Thus we can summarize the models under study as follows:

(1) *Bibliometrics (lineage or cladistic concept).* We performed bibliometric trend analysis based upon the citation network of documents. The hypothesis behind this was that the network (or, more precisely, its inverse relation, see Section 5 below) conveys a thematic descendancy, or ancestor–descendant relation between publications unfolding along the relevant time scale, thus expressing the evolution or dynamics of the field under study.

(2) *Content analysis (morphological or similarity concept).* This analysis was based on the textual relations of documents. The relation was defined by keyword-based similarity between documents, constituting a proximity network upon them. The hypothesis behind was that the time dynamics of this network mirrors the emergence and development of thematic clusters or foci constituting the changing structure of a research field.

(3) *Social (author) network analysis (biological species concept).* This type of analysis was based upon the social view of research fields. The best candidate for the definitive relation was co-authorship, used in grouping the literature representing the field. The dynamics measured here was the birth and change of author communities.

## *4. Materials*

For conducting a multidimensional trend analysis as described above, we have picked an influential, multidisciplinary research topic that ranges from philosophy to psychology to artificial intelligence and sociology, namely, the scholarly discourse on *intentionality*. Intentionality is both a long-standing and a multidisciplinary topic with many known sub-topics descended from a historical, „common ancestor". A diffusion of this discourse through disciplinary boundaries also makes it ideal for a study aiming to compare topical speciation in different models of the evolution of science.

Bibliogaphic data on the scholarly discourse was retreived from three databases available through the Web of Science™: to cover the multidisciplinary scope of the topic, we consulted the *Science Citation Index* (SCI-EXPADED), the *Social Science Citation Index* (SSCI) and the *Arts and Humanities Citation Index* (A&HCI). Records of publications containing „intentionality" in either the *Subject* or *Title* fields were harvested from these sources. The time coverage, i.e. the year of publication was set to the range 1975–2009 (inclusive). This procedure resulted in a set of 1934 publications for 35 years.

In order to test the dynamics extracted by the three aforementioned models upon this dataset, we partitioned the 35-year corpus by the year of publication into 7 consecutive and equal time intervals. These 5-year periods provided the basis for a longitudinal comparison of the three models (see the description of network construction below). As can be seen in Fig. 1., the distribution of documents over these periods was quite unequal, and reveals an exponential trend: the first time window represented less than 100, while the last covered more than 600 publications. This also means that the comparison of the models, applied in a step-by-step fashion to each period, was based on a small sample size in the first few steps. Another option would have been, as opposed to the applied time-slicing method, a slicing of the document set into partitions with equal sizes, thus containing a similar amount of documents each, which is a common practice in scientometrics. Yet time-slicing seemed more suitable for our purposes than size-slicing, for at least two, related reasons: (1) our main aim was to compare the different patterns of timely dynamics, and (2) size-based slicing would have aggregated large periods of thematic developments, therefore supressing the dynamics under study.

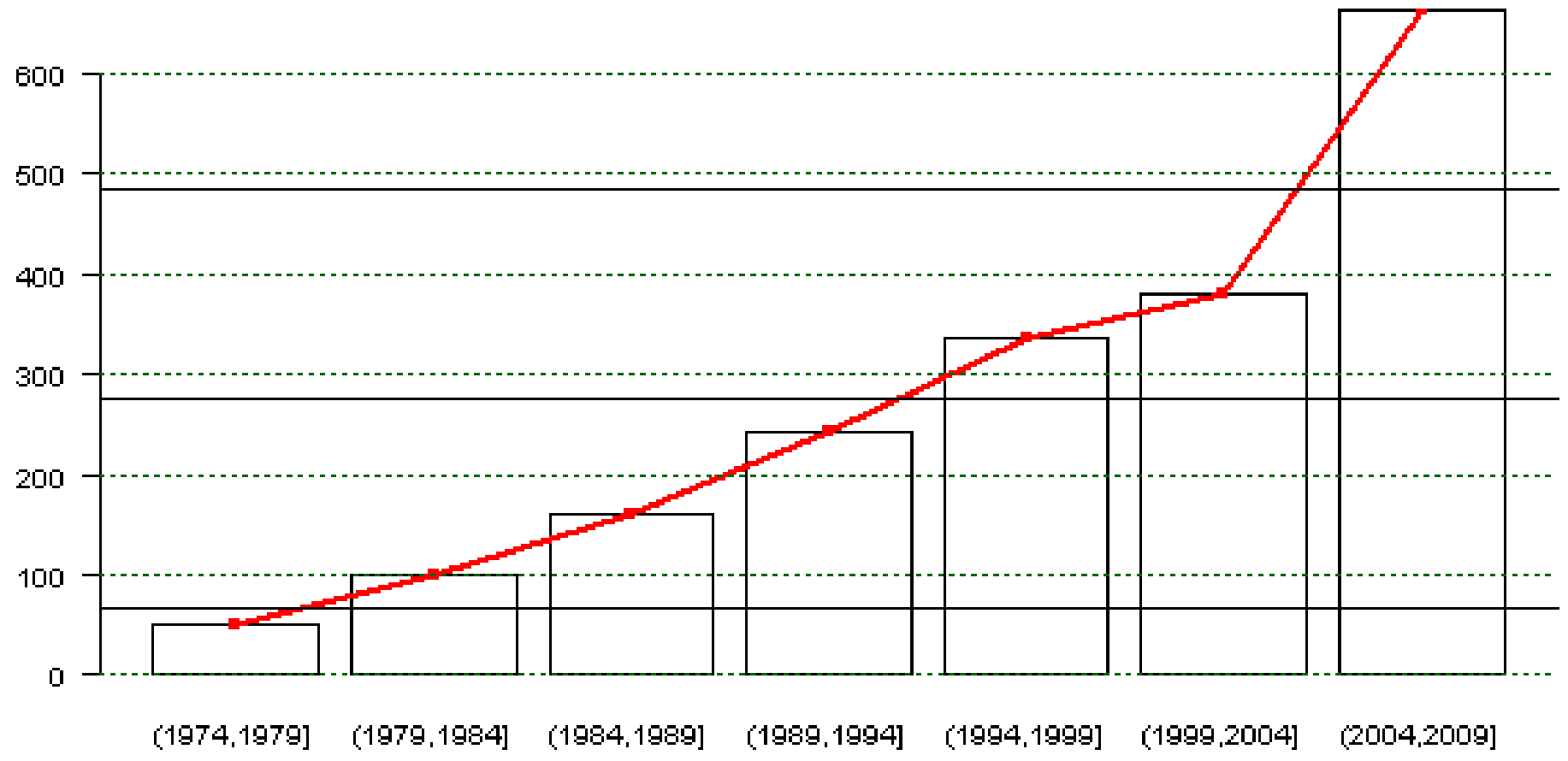

Fig. 1. Distribution of the „intentionality" corpus over seven time periods

### *5. Dynamic network models of scientific species*

As the next step, we built the three models (bibliometric, content-analytic, and co-authorship i.e. lineage, morphological and „biological species" related models) for the evolution of the concept of „intentionality" in the form of *dynamic networks*. The construction of each of the models consisted of the following blocks:

(1) extracting the network of documents for every time slice, thereby jointly constituting a time series of networks, or a dynamic network, and
(2) applying a suitable method of network community detection for the individual networks.

The second step serves as the basis for trend analysis in the narrow sense: document clusters resulting from the various community detection methods are conceptualized as sub-topics or sub-discourses of the subject matter in each time period (ie. in each network) – or, from the biological perspective, conceptualized as synchronous species. Community detection methods were chosen to implement the species definitions associated with the above discussed bibliometric, content analytic and social models, respectively. In all cases, network construction and analysis were conducted using the R-based packages of the TexTrend toolkit (http://www.textrend.org). Details of network construction and the clustering methods for the three models are summarized here.

*(1) Bibliometric model (lineage concept).* To represent the descendancy relationship in the corpus, for the bibliometric case we made a coupling between the publications (source documents) and the references of publications, in each period. After the normalization of source documents, representing those by the same syntax as that of the reference set, we extracted the citation networks in each case. This operation resulted in a series of directed (unweighted) graphs: in each graph with a time parameter *(t-t')*, the nodes represent source documents published between the years *t* and *t'*, as well as documents cited by any of these sources. Edges of such graphs show the *x– is citing– y* relation between publicatons. Since we were interested in a kind of intellectual descendancy in the corpus, therefore, as the key step, we generated a new network series by inverting this relation (reversing the direction of edges): the resulting new networks can be conceived as expressing the *x–is descended from–y* relation.

*(2) Content analysis model (morphological concept).* The model capturing content-based relationships between documents was designed to form a close analogy with the „statistical concept of species", i.e. the species concept of numerical taxonomy. This concept, often referred to as the phenetic species concept, is best known (in its original form) for utilizing as many morphological features as possible to statistically estimate the similarity of organisms, and thereby establish similarity clusters as candidates for real species. In our case, organisms were replaced by documents, and morphological features by metadata of documents. Phenetic similarity was instantiated by the proximity of articles in terms of their content-based descriptors: pre-defined keywords provided in the database.

To map the similarity clusters for each studied period, a new set of document networks were constructed. Each graph, as in the previous case, corresponded to a time window in the consecutive series of intervals, containing the source documents published within that period. Graphs were based on the term–document (incidence) matrices of these periods, the terms being keywords occuring in the document set. Edges in the graphs represented the cosine similarity of document vectors: an edge weight, therefore, conveyed the degree of sharing the same set of features (keywords) by two articles. This procedure resulted in a textual similarity network, changing through time, and composed of the elements of the document corpus.

For a more comprehensive insight, we utilized two types of keywords describing the topic of individual publications. The ISI databases contain at least two different keyword-like descriptor fields provided for the summarization and categorization of documents. The first we used were author-generated keywords (DE), and the second was indexer-generated keywords (ID). As the terms suggest, DE-keywords are freely selected by the author, and give a more detailed and specific characterization of the content, while ID-keywords, being consolidated to a more-or-less standardized terminology, are less subjective and more suitable for the purposes of categorization (however, both methods are prone to the indexer effect). In our model, using both types of descriptors translated into building two distinct series of publication nets, one for capturing document similarities in terms of DEs, and the other for IDs. From the perspective of the species concept, these two relations are two different candidates for defining scientific species: author- and indexer- generated keywords count as two, (conceptually) non-overlapping classes of morphological features. Whether „species” based on DEs are identical with those based on IDs is exactly the type of empirical question that our experiment was designed to answer (see Section 7).

(3) *Author network model (biological species concept).* The third model of scientific species was inspired by the most prominent concept of species, the so-called „biological species concept”. Instead of descendancy or phenetic similarity, clusters under the biological species concept are defined by the capability to interbreed: that is, by a compatibility relation between organisms. Finding the counterpart of this relation in the bibliographic domain – if any – is a non-trivial task: however, there is a dimension of structural scientometrics that quite naturally lends itself to embrace the notion of „interbreeding”. Co-authorship patterns are a traditional target of science mapping: and co-authorship can be viewed as a kind of „intellectual interbreeding”, resulting in joint publications. Though the analogy ends at this point, we can turn this metaphor into a model, sensible to scientometrics, where the social relation of co-authoring an article counts as the criterion for defining sets of related documents. That is, under this approach, documents that share authors are conceived as being related.

The network model implementing these ideas took the following form: for each period, the author–document matrix was computed, based on which a proximity measure between documents were defined. The proximity of two documents, in this case, was simply formalized as the number of common authors. This yielded a series of document networks for the consecutive time intervals. Each network was composed of source documents published in the corresponding period; the edges indicated that the nodes (publications) connected were sharing some authors (with edge weights = number of authors shared). Note, that this structure would count as the inverse of traditional author-networks. Network architecture, in this case, also conveys author community patterns; however, by having documents as the unit of analysis (instead of authors), the comparison of the „biological" model with the previous two models (the lineage model and the morphological model) has been made straightforward.

## *6. Species definitions*

Given the three models defined above, the next step of our experiment was to detect the „species" in each case, that is, to circumscribe cohesive groups of documents based on the bibliometric, the content analytic, and the author network model, respectively. To stress the biological analogy, the methods applied for this task were also selected to be in accord with the corresponding species concepts (the lineage, the phenetic and the biological species concept). A widely held view in theoretical biology (and in the philosophy of biology) is that any modern definition of species might be decomposed into two parts: (1) the choice of the relation (between organisms), on which the concept is based, and (2) the method of partitioning the relation into species.

So far, applying these ideas to the bibliometric domain, we have identified the three definitive relations, as that of descendancy (*x-is-descended-from-y*), compatibility *(x-shares-authors-with-y)*, phenetic similarity *(x-shares-keywords-with-y)*. Those, who propose any of the first two relations in biology, also propose a basically identical method for partitioning, while proponents of the phenetic approach should rely on a different way of slicing up their network.

For both the lineage/descendancy and the biological/compatibility species concepts, defenders argue that partitions are naturally yielded by these relations, i.e. that no additional criterion is needed: for instance, it can be aruged that (maximal) groups of potentially interbreeding organisms are connected by this very relation, and isolated from other such groups. The case of descendancy is a bit more complicated, since, by choosing a large enough time window, every organism is connected to every other through common descent (the „tree of life"). However, using subsequent periods on the evolutionary timescale quite naturally cuts this tree into synchronic lineages, that is, into distinct groups co-existing in the same time window, within which organisms are interlinked by descendancy, but between which there is no connection *in that time interval.* As opposed to descendancy- or compatibility-based species, phenetic species are grounded in a statistical relation: the degree of similarity. Therefore, pheneticists should incorporate a more-or-less tentative criterion into their definition, the threshold on similarity, above which two organisms are to be sorted into the same species. Though this threshold or rule is usually an embedded feature of some sophisticated statistical clustering or classificatory procedure, it still represents a less „natural" constraint on the boundaries of species than do the previous two approaches.

We have implemented the following procedures to detect species in our bibliographic networks:

*(1) Bibliometric model – (2) Author network model.* In both of these cases, we let each network to „slice itself up", without imposing any additional rule on partitioning or community detection, just as the corresponding views on the lineage concept and the compatibility (interbreeding) concept dictate. In practice, this idea is equvivalent to taking the maximal connected components of a network as separate species. A maximal connected component is a set of nodes in the network, any two members of which are connected directly or indirectly (that is, through a series of edges), but each member of which is separated from all other nodes (cannot be reached from outside the component). In this way,

(1) bibliometric species were defined as the components of the descendancy network of documents in each time period under study;
(2) author-network species were defined as the components of the author-similarity network of documents in each time period under study.

_(3) Content analysis model._ For the counterpart of the phenetic concept, a proper community detection method was needed, to circumscribe species in each dense document similarity network (in other words, some threshold was necessary to be set), in the same way as the multidimensional feature space of organisms is subjected to some partitioning procedure in numerical taxonomy when seeking species in the dataset. To avoid selecting some arbitrary threshold, we utilized an iterative method of community detection, that was designed to slice up dense networks in „the best natural way”. We used the Walktrap Community Findig (WCF) algorithm (Pons & Latapy 2005) as implemented in the iGraph R package by Pons & Csardi (no date) and Csardi & Nepusz (2006), that attempts to find dense subgraphs within a network by random walks. The underlying idea for this algorithm is that short random walks with the probabilities determined by the edge weights are likely to circumscribe a community in the sense of being a set of densely and strongly connected nodes.

The WCF algorithm works in an agglomerative fashion, starting with the strongest communities and merging the closest ones in consecutive steps until the whole network is reconstructed. For optimization we used the modularity function of Newman and Girvan (Girvan & Newman, 2002; Girvan & Newman, 2004; Newman, 2006):

$$Q = \frac{1}{2m}\sum_{i,j} A_{ij} - \frac{k_i k_j}{2m}\delta(c_i, c_j),$$

where $m$ is the number of edges, $A_{i,j}$ is the corresponding element (weight) of the similarity matrix, $k_i$ and $k_j$ are the degrees of the corresponding nodes, $c_i$ and $c_j$ are the cluster indices the two node belongs to, respectively. $\delta(c_i,c_j)$ is a function that equals to 1 where both nodes are the same clusters ($c_i = c_j$), and 0 otherwise. Informally speaking, the function measures how “modular” is a given network under a certain partition of its nodes (community structure), in how separated the different node types (clusters) from another. Using this measure as the object function to be maximized, we selected the community structure of highest modularity. To put the same thing in yet another way, species in the third model were those groups of thematically similar documents that jointly represented the most modular decomposition of the network into adjacent communities. (This procedure was applied on both series of keyword-similarity networks, that is, on DE-networks and ID-networks, separately.)

## *7. Comparisons of the three models*

The central aim of our experiment was to compare the three models of scientific species. Recall that we were mainly interested in question (2) within a series of questions formulated in Section 2 concerning the relationship of the bibliometric, content analytic and social network-based identifications of research trends. Namely, we are focusing on the question about the extent to which the three dynamic models converge, that is, the degree to which they pick up the same research trends, and lead to similar conclusions as to the dynamics of particular scientific discourses. „Dynamics" is a key term here, since we were to capture not only the general agreement between three kinds of structuring our corpus as a whole, but also to detect the overlap in the formation of these three structures through time. This was the reason to slice up the corpus into consecutive time periods, with documents belonging to them. Using these intervals, we conducted a step-by-step comparison of the three models, contrasting their structures within in each time window.

To implement these comparisons, we chose the following method. The three species detection procedures described in the previous section provided us with three different partitions (community structures) of the same set of documents for each time period. At this point, we also gain an explanation of why we use documents as the unit of analysis within each model (instead of, for example, authors or keywords in the social network and the morphological model, respectively): in this way, the methods are directly comparable, since the application of the lineage, the morphological and the compatibility method uniformly result in classifications of (the same set of) documents.

To assess the agreement between the three groupings of documents and their time evolution, we used a well-known tool invented to measure the similarity of different clusterings: the Rand index. Given two partitions $P$ and $Q$ of the same set of elements, one can define the following types of relations between any two members $i$ and $j$ of the set (where $i$ an $j$ are not identical):

- $i$ and $j$ are classified together according to both partitions $P$ and $Q$. Let the number of such pairs denoted by $SS$ (similar–similar).
- $i$ and $j$ are classified differently, into separate classes according to both partitions $P$ and $Q$. Let the number of such pairs denoted by $DD$ (different–different).

- *i* and *j* are classified together according *P*, but differently by *Q*. Let the number of such pairs denoted by *SD* (similar–different).
- *i* and *j* are classified differently according *P*, but together by *Q*. Let the number of such pairs denoted by *DS* (different–similar).

Based on these relations, the Rand index (Rand 1971) is calculated as follows:

$$R = \frac{(SS + DD)}{M}$$

Here *M* refers to the number of all possible pairs of elements in the set, that is, $M = SS+DD+SD+DS$. In other words, the Rand index measures the ratio of pairs treated the same way in both partitions to the amount of all possible pairs.

Using this approach, we calculated pairwise Rand indices for each combinaton of models for every time period (we implemented the calculations utilizing the R statistical software: R Development Core Team 2009). That is, we generated all the specific Rand indices indicating the overlap between the bibliometric grouping, the content analytic grouping and the author network grouping for every time interval. Such an exercise would involve 6×7 = 42 comparisons, since we had six pairwise combinations of the four models (including both types of keyword-networks, DE- and ID-groupings, remember that the morphological concept actually covered these two models), and seven time intervals. However, this amount reduced, in practice, to 5×4+7 = 27 figures, due to the fact that the morphological models covered only the last four time intervals in the corpus, therefore, in all the five combinations with the (two) morphological structures only four time periods could be accounted for. [1]

Results from these comparisons are depicted in Figure 2. In each plot, a specific comparison of two groupings can be observed together with the evolution of their overlap (in most cases, only the last four periods, due to the above mentioned reasons). Figures on bars stand for the value of the Rand index for the corresponding time window. It is quite straightforward from these series of diagrams that the similarity of any two partitions or species structure shows little dynamics, in terms of the Rand index (but see different observations for altering the applied measure below). The overlap of any two structures seems rather stable at some level, except for the last comparison, the author-network and the bibliometric model (Fig. 2f). The agreement between these community structures decreases from R=0.97 (second period) to R=0.58 (last period). A smaller fluctuation also shows itself in the relation of content analytic and bibliometric species (Fig. 2c).

[1] This latter condition was a drawback of using bibliographic data retreived from the ISI databases: documents published before the fourth time period (1999, 2004] were not systematically provided with either type of keyword (DE or ID) in the database, so these parts of the corpus could not be subjected to the content analytic model.

The lack of a heavy dynamics thus makes it easier to evaluate the behaviour of individual models againts each other. As might be expected, the two morphological models, that is, species from DE-keyword similarities contrasted with species from ID-k eyword similarities, are in a considerable agreement (R values range from 0.81 to 0.9, Fig 2a). Somewhat unexpected is the result that author-network species outperform this extent of overlap in every combination with morphological species (Fig. 2b–2c): both with ID-clusters and DE-clusters their agreement is uniformly above R=0.9. On the contrary, bibliometric species (the lineage model) exhibit a moderate convergence to any of the morphological communities (Fig 2c, Fig 2e), with R values around 0.5. For many periods (timeslices 1–6), there is a closer match between bibliometric and author network species (Fig 2f), with a decreasing slope over the timescale.

What causes the observed surprising patterns of „co-evolution”? A closer look on the three time series of (partitioned) networks explains a great deal of what is going on behind the formation of Rand values in the comparisons above. Visualizations of the bibliometric/lineage graphs, the morphological graphs and the author-network graphs are made available in the Appendix of this paper. A key observation concerning the architecture, and, most importantly, the community structure of these networks is that, in comparison, the size of „species”, and the distribution of documents among species differs heavily when we switch from bibliometric to morphological to author network species.

In particular, for bibliometric species (Fig. A1), we can divide the timescale into two sections: in the first section (periods 1-3), a few small clusters emerge, and most documents are „singletons”, that is, form a cluster on their own. In the second section (periods 4–7) a single large cluster and some smaller ones is characteristic of the picture (with a „supercluster” unifying the vast majority of documents in period 7), the remaining articles acting as singletons, again. In the case of both kinds of morphological species (Fig A2–3)., a much more even distribution between species is given. ID-clusters and DE-clusters are relatively small communities with similar sizes. Furthermore, in the morphological cases no singletons left after species formation: each document belongs to a small but never single-membered group. Still different is the behavior of the author-network model: In these graphs (Fig. A4), two-membered species seem to be the rule, within which some more populated and more complicated clusters show themselves, mainly in the later periods (in period 7, a partially connected larger subgraph also emerges).

Given these differences in the resulting structures, it is reasonable to reconsider the performance of the Rand index in their comparison. Note, that the R index sums up those documents that are categorized similarly (SS) or *differently* (DD) by two models. Now that we have seen our networks, this feature directly explains the relatively bad performance of the bibliometric concept in almost each combination (but mainly with the morphological models). On one hand, it contains a huge number of singletons, „categorized" differently, that are distributed in small clusters in the morphological model, categorized together (DS-type pairs). On the other hand, it also contains a large cluster with many unified documents that are also partitioned apart in those small clusters in the morphological cases (SD-type pairs). Therefore, SS and DD-type pairs are relatively rare in these timeslices. The very same feature of the Rand index, however, accounts for the contribution of the author-network model in these comparisons. Clusters or species of „common authorship" are very small, so the chance of being sorted to disjunct classes is rather high; when compared to the more extensive but still small morphological clusters, one might expect a lot of differently categorized pairs as a result. Indeed, when directly inspecting the factors of the Rand values in these couplings, what is striking is a very high amount of DD-pairs, in relation to any other types (SS, DS, SD). So, in this case, the almost maximal R-indices are attributable to differently categorized documents, instead of those sorted similarly by the two models. A mixture of these two cases (biblometric species/author network species vs. morphological ones) is present in the bibliometrics vs. author network case. In the first periods, differently categorized pairs (DD) contribute to the high R values observed, due to many singletons and a few middle-sized clusters on the bibliometric side, together with the two-membered clusters on the author-network side. In later periods, however, the increase on the barplot indicates that the bibliometric structure gradually generates large species (i.e. many similar documents), the other side remaining quite scattered (non-similar documents). In other words, the number of SD pairs explodes at the expense of SS and DD pairs.

Fig 2. The evolution of overlap between the bibliometric, the content analytic and the author-network „species" or sub-discourses, expressed by the Rand index between these groupings in each respective time period.

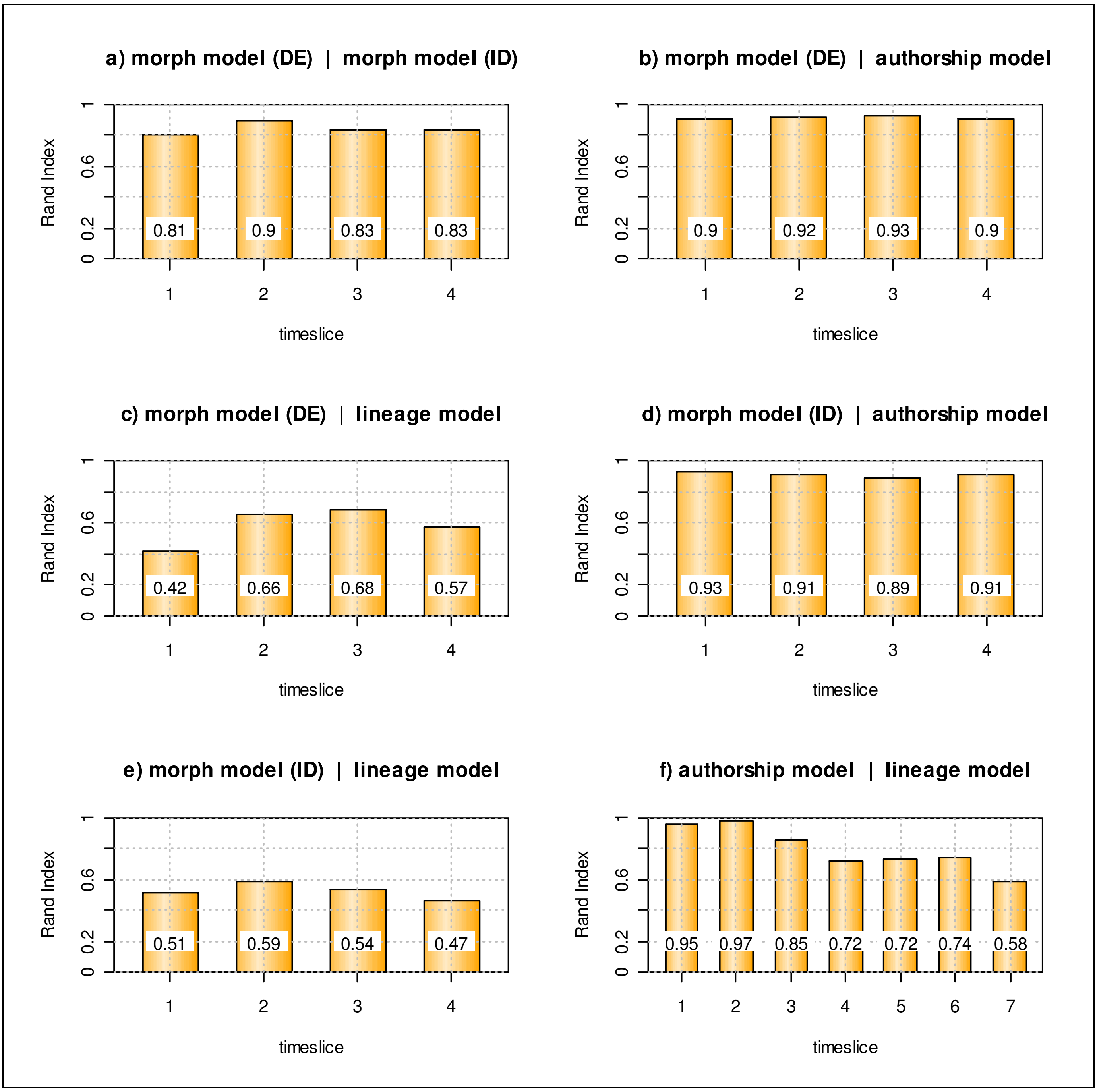

According to the discussion above, the results of our comparative experiment are primarily shaped by two factors: (1) the characteristic community structure of networks under each „species concept", and (2) the terms included in the Rand Index. In relation to the latter, one might argue that the series of high values produced by this formula actually fail to capture the proper relationship of the groupings included, based on the observation that those peaks are accounted for by differently categorized pairs (DD). Since we are interested in the question of whether a species in one of these groupings matches another species in the other grouping, it would be more appropriate to take into account pairs that are of the same species in at least one of these classifications. The argument can even be sharpened by noting that counting singletons or even two-membered classes as an increment of the similarity in species structure (as an increment of differently categorized pairs) is far from reasonable, since these are not proper communities in the general sense. Again, singletons could be excluded from comparisons by focusing only on documents that are in a proper cluster in at least one case, in other words, considered similar under some model of interest.

To refine our assessment, and also to overcome the potentially misleading import of the Rand Index, we selected a somewhat different measure to implement a second comparison incorporating the lesson of the previous argument. This measure was the Jaccard index, defined here as follows:

$$J = \frac{SS}{SS + SD + DS}$$

with the previously introduced notation. As can be seen, the measure provides the amount of (pairs of) documents belonging to the same species under both classifications (SS) as the proportion of those pairs that belong to the same species under either one of the two classifications (SS, SD, DS). This measure, therefore, conveys the notion of two models „ruining" or matching each other's communities. We reiterated our calculations using the Jaccard index as a replacement for the Rand index, which led to the interesting results depicted in Figure 3.

Fig 3. The evolution of overlap between the bibliometric, the content analytic and the author-network „species" or sub-discourses, expressed by the Jaccard index between these groupings in each time period.

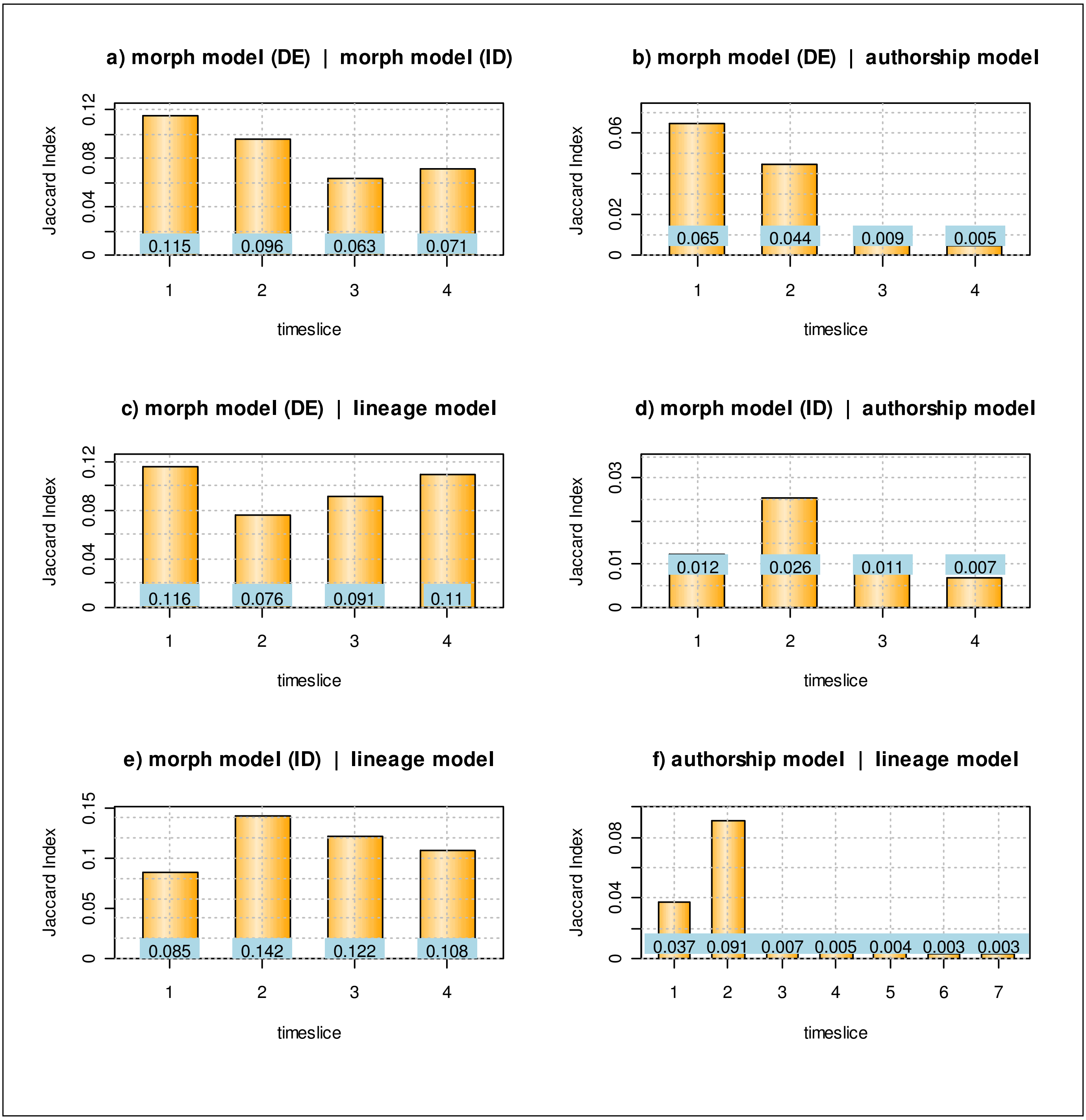

Combinations of models, evaluated by the Jaccard index, occur in the same order in Fig. 3 as in the Rand index Figure. In general, two instant changes are apparent as compared to the previous series of barplots: (1) values of agreement range within a much lower scale than before (for the sake of visibility, scales are adjusted to the order of magnitude of the resulting values in each plot), and (2) the co-development of the structures shows more variability and dynamics than in the case of R-values.

Most importantly, the match between species structures also exhibits a new arrangement in terms of the J index. The overlap between the two morphological models (DE-species, ID-species, Fig 3a), as the sole exception, remained relatively intact (though reduced in absolute figures), but in this case, the two classifications diverge along the timescale (from J=0.115 to R=0.071). In all other cases, however, we experience the opposite of the tendenies presented above. For now, the bibliometric/lineage model is the best matching species set in most combinations (with the morphological counterparts). Interestingly, apart from their overall match, the dynamics of the lineage model as contrasted to ID-species and DE-species, respectively, seems to be the inverse of each other. DE- and bibliometric species start with a higher agreement (first period), and then, after a sudden fall there is a gradual increment period-by-period, while eventually the inital value is being reached again. (Fig. 3c). On the contrary, with ID-species, after a moderate start a sudden peak occurs (second period), followed by a moderate negative slope till the last period (Fig. 3e). As to the contribution of author-network species in this collection, the match with any other model is, in most cases, an order of magnitude lower than the agreement between other species structures: especially so in relation to bibliometric species in later periods (Fig 3f). A striking outlier in the latter time series is the second period, whereby a sudden peak occurs: in this particular interval, a few small clusters rule both patterns, lineage species being yet „underdeveloped", and the author-network model moderately scattered. In later timeslices, however, the two patterns heavily diverge, as the author-network model is becoming mostly unconnected, while the bibliometric concept is connecting most documents in an evergrowing single species. With morphological species the situation is not so dramatic: relative to the other cases, there is a considerable overlap with DE-communities in the first available time window (that is, the fourth period), which also decreases an order of magnitude through time (Fig 3b). At the same time, the type of keywords used in detecting communities seems to make a difference, since ID-species do not fit with author-network species as much as DE-species do: with little variance, the match is about 0.01 for each timeslice.

In sum, by the Jaccard index and, therefore, by a more rigorous notion of species structure similarity, the closer kinship of the morphological/content analytic and the bibliometric/lineage concept is favoured over the other combinations.

## *8. Conclusion*

In this study we addressed the timely question of contrasting the methods in science mapping utilizing a context, where similar problems have long arisen: the problem of competing species concepts characteristic of biological systematics. One of our central claims was that these scientometric methods can be reconstructed in terms of species mapping under rival conceptions of biological species. We argued that formalizing the major methodologies introduced for the ideal detection of scientific specialities and trends in terms of corresponding species definitions would support the comparison of these methodologies. Most importantly, this kind of comparison is best designed to cope with the dynamic aspect of the formation of scientific subjects and communities.

Therefore, after the retrieval of a bibliographic corpus covering the discussion on a sample concept, *intentionality*, a topic quite interdisciplinary in nature, we defined three models instantiating the citation-based, the keyword-based and the author-based mappings of scientific trends to compare their performance in the trend analysis of this subject. These models were motivated by, and analogous to the so-called lineage concept, morphological concept and biological concept of species, respectively. The overlap of the structures of this literature as identified by the three models were measured by the Rand index and the Jaccard index in seven consecutive time periods to expose the dynamics of the co-developments in each pair of models. An important difference in relation to previous science mapping techniques was that, in each case, documents served as the unit of analysis, which made the resulting groupings directly comparable.

The analysis of the three groupings against each other, each embedded into some major scientometric tradition, revealed considerable differences as to the structure of the corresponding dynamic networks. These differences heavily contribute to the result that, evaluated by the Rand index and the Jaccard index, respectively, little convergence between species structures can be observed: it turned out that the Rand index pretty much outlines the differing nature of these concepts, while the Jaccard index shows the moderate agreements between specific models.

Quantitatively speaking, pairwise comparisons (with the use of the Jaccard inex) confirm the relatively close relationship between content analytic/morphological and bibliometric/lineage document clusters (as contrasted with other combinations). This relation is of particlar interest for scientometrics, since the keyword-based method, and the reference-based method are two competing paradigms of knowledge mapping. References are usually considered to be as providing the intellectual basis of an article, while keywords are conceived of as positioning the corresponding paper in the thematic space of some speciality or research area. Both kinds of metadata, therefore, thought to be serving as proxies for the empirical delineation of some direction of research. On the other hand, the individual study of both of these methods raised several problems as to the capacity of grasping real trends, research fronts and communities (cf. King 1987, Braam–Moed–Raan 1991a,b). Our study points to the direction that documents related along citation lineages (and by co-citations) tend to exhibit a common dictionary, to some extent, or that ineage-based and content-based approaches show some convergence. This supports the view that the two ways of mapping science approximates valid specialities or communities of research.

Although, based upon the presented experiment, we cannot generalize our claims, some interesting furhter hypotheses might also be formulated. Such a hypothesis would be that the dynamics shown by the co-variation of the two content-based and the lineage-based species is an indication of a periodic nature of this development. Recall the observation that, in both cases (though in different ways) the process „starts over again", that is, after an initial state a sudden change occurs, followed by a gradual monotonic one, through which the initial state is more-or-less re-instated. We might interpret these patterns that, in the case of author-generated keywords, after a paradigm change, periods of canonization are exhibit themselves, in which a common lexicon is being formed along descendancy lines, or a paradigm is gradually being constructed. The reverse appears in the case of indexer-generated keywords, suggesting that external lexicons gradually loose their capability to cover developments in the field, but are updated time to time in order to catch up with the evolution of science.

In sum, we might claim that the comparative analysis of traditional science mapping practices with the conceptual aid of the species problem is in good support of exploring th meaning and capacity of these traditions to outline scientific trends. The present paper is best considered a pilot study in this direction: re-iterations of this approach on other, and more extensive samples of different subject matters are planned in order to test the hypotheses introduced in our exploratory work.

### *Acknowledgment*

We acknowledge the financial support from the Future and Emerging Technologies (FET) programme within the Seventh Framework Programme for Research of the European Commission, under the FET- Open grant number 233847 (Dynanets project, www.dynanets.org).

***Appendix***

A1. Development of the bibliometric/lineage clusters from period 1 to 7 (graphs contain both the set of source documents and that of cited documents).

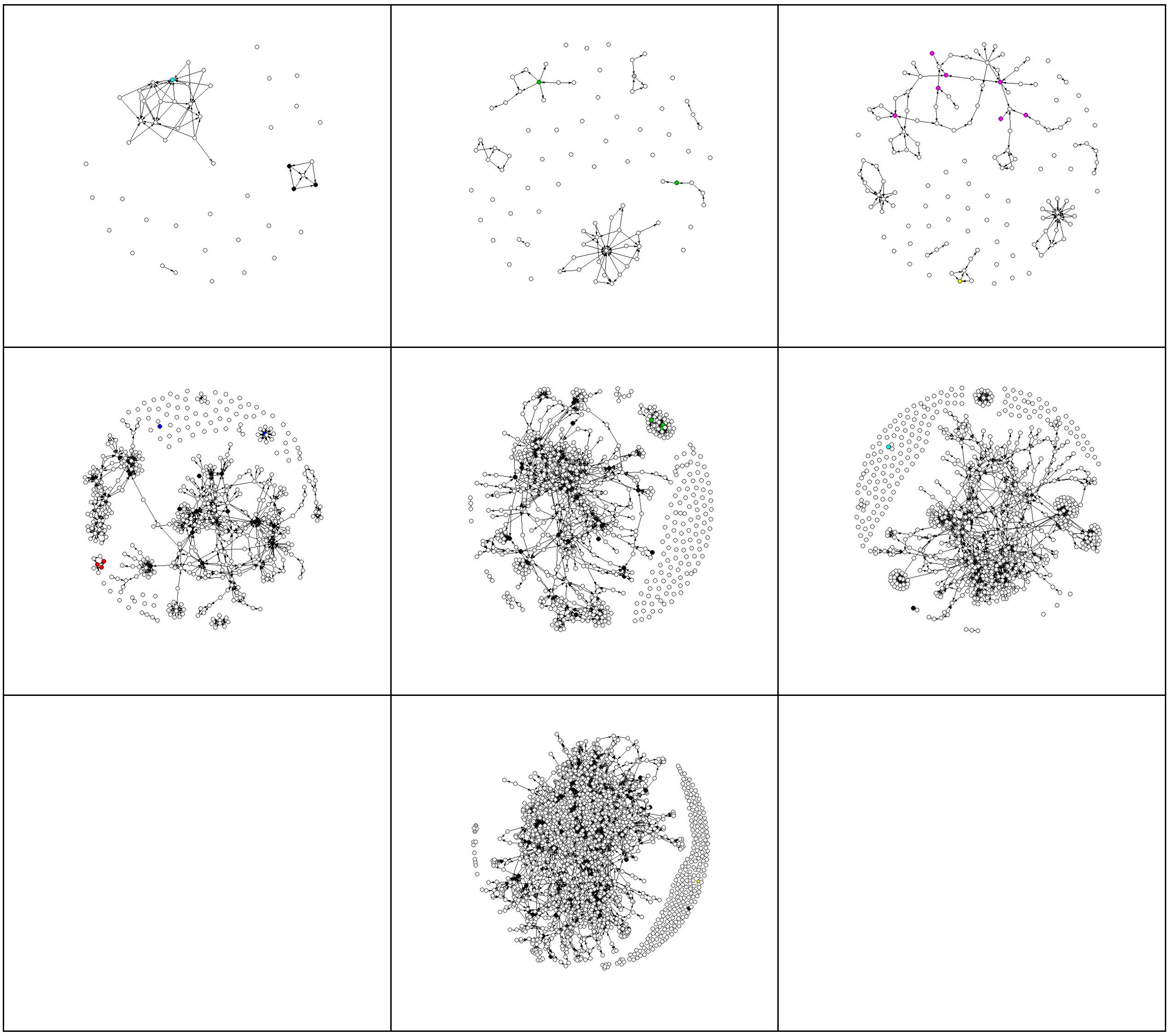

A2. Development of the DE-similarity clusters from period 4 to 7.

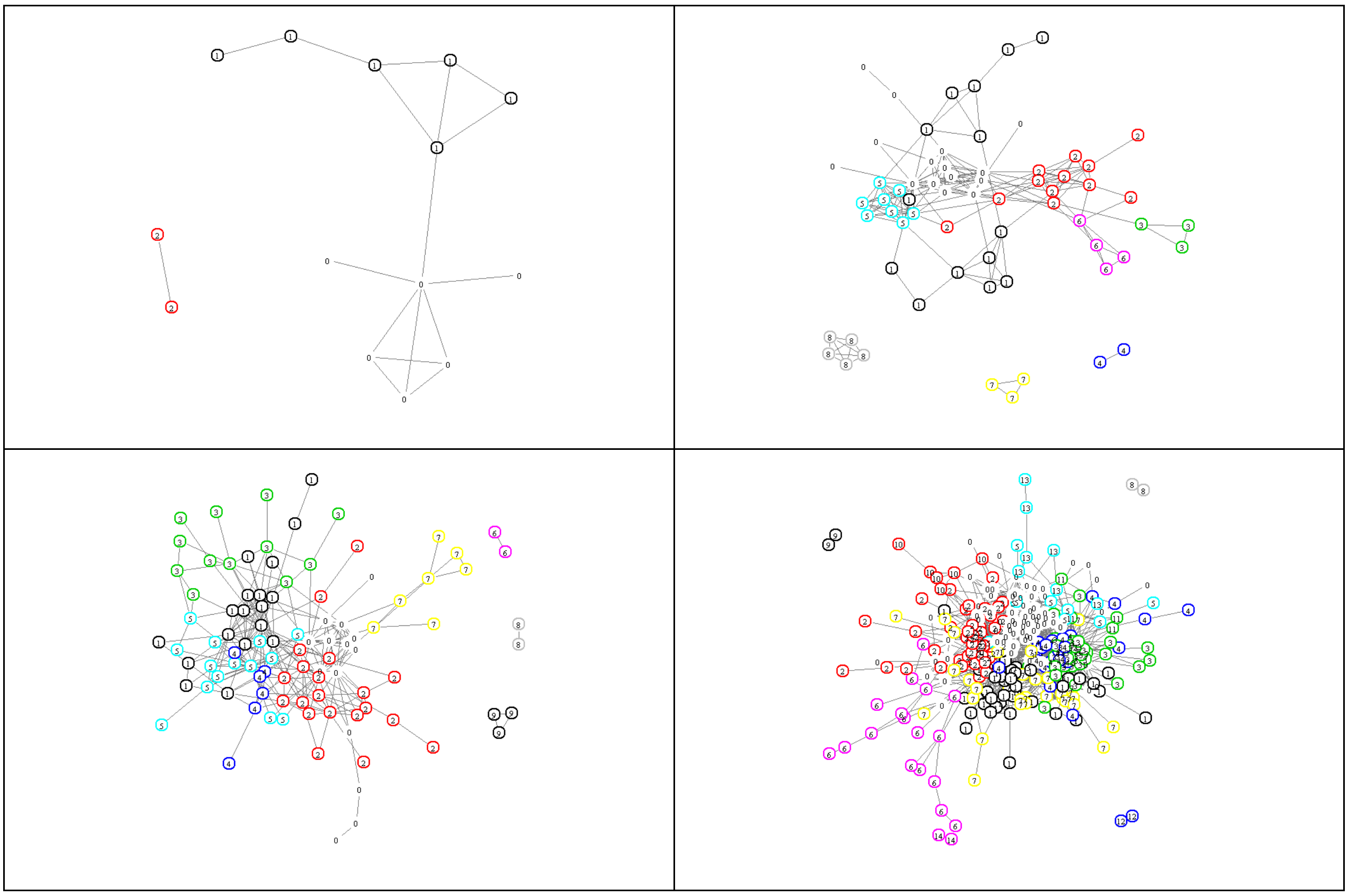

A3. Development of the ID-similarity clusters from period 4 to 7.

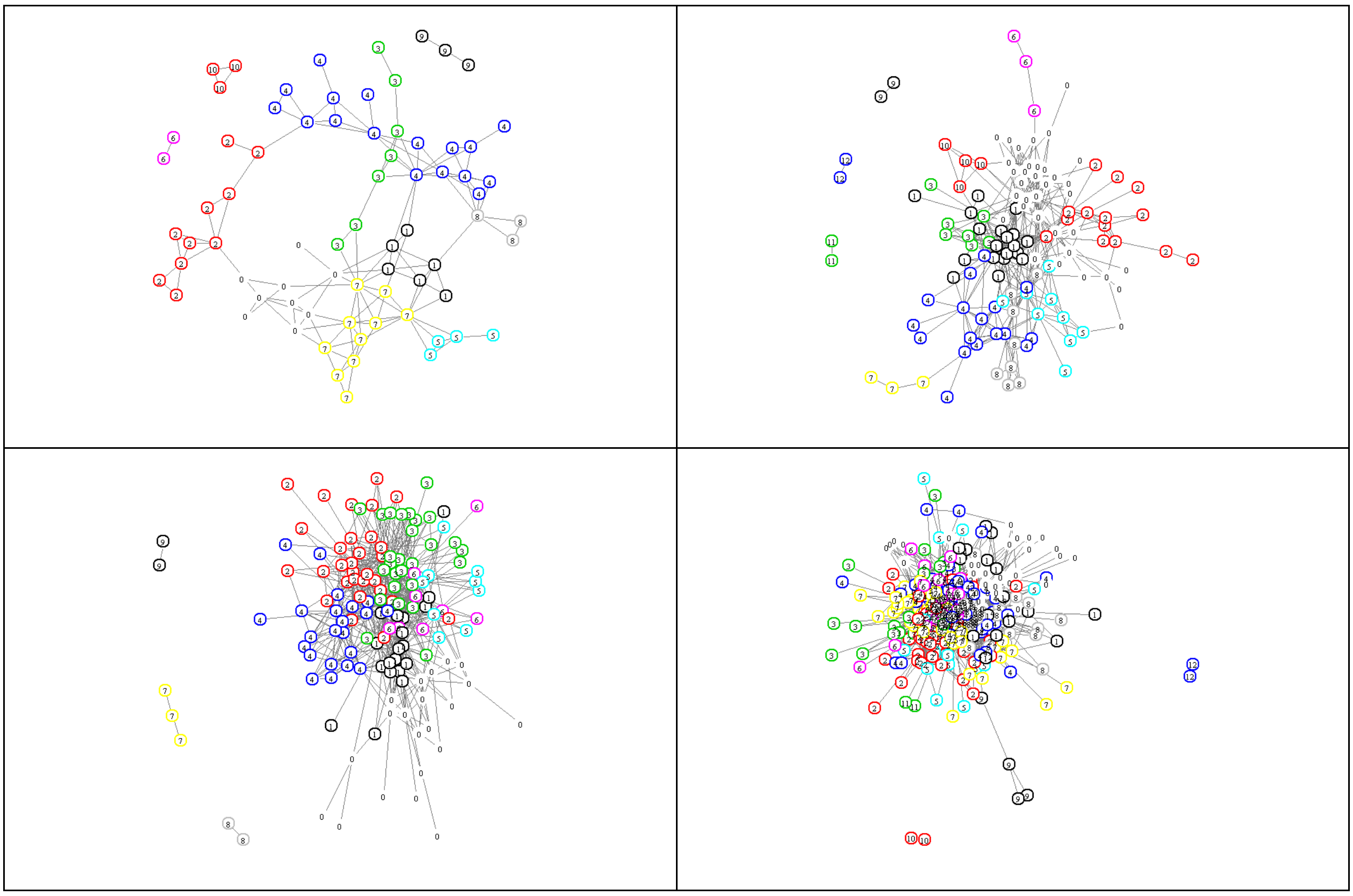

A4. Development of the author similarity clusters from period 1 to 7 (singletons omitted from the graphs.).

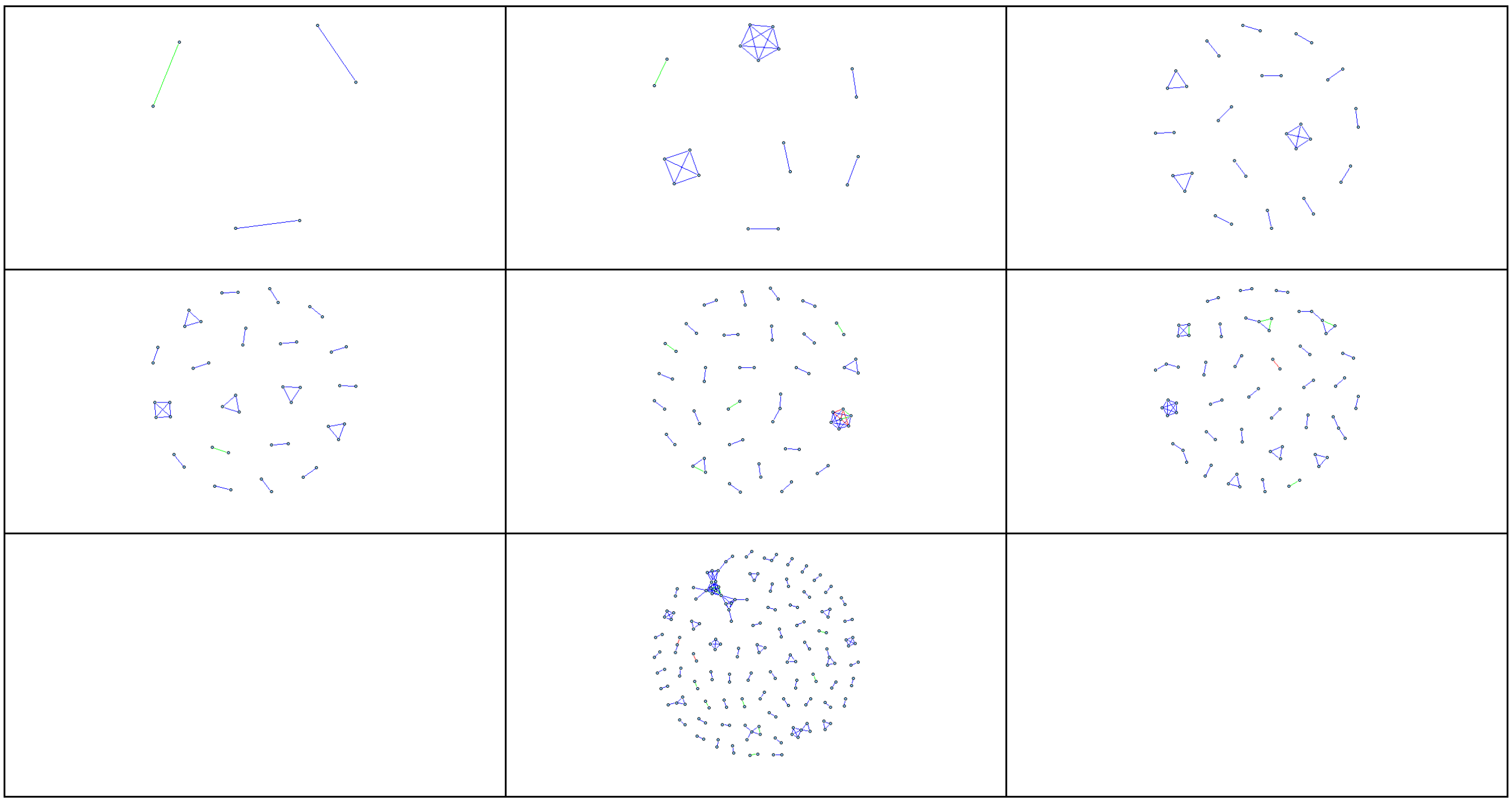